\documentclass[11pt]{article}
 \renewcommand{\baselinestretch}{1.03}
\newcommand{\mra}{\>\!\!\bigr\>}
\newcommand{\mla}{\bigr\<\!\!\<}

\newcommand{\be}{\begin{equation}}
\newcommand{\ee}{\end{equation}}
\usepackage{amsmath}
\usepackage{amssymb}
\usepackage{latexsym}
\usepackage{epsfig}
\usepackage{cite}

\providecommand{\href}[2]{#2}

\def\<{\langle}
\def\>{\rangle}

\newcommand{\del}{\partial}

 \newcommand{\myBoxed}[1]{\boxed{\phantom{\biggl(} #1 ~}}

\begin{document}

\begin{titlepage}
\rightline{\tt PUPT-2350}
\rightline{\tt UT-Komaba/10-6}
\rightline{\tt IFT-UAM/CSIC-10-51}
\begin{center}
\vskip .6cm

{\Large \bf {Solutions from boundary condition changing operators\\[1ex]
in open string field theory}}
\vskip 1cm
{\large {Michael Kiermaier${}^{1}$,~ Yuji Okawa${}^{2}$ ~and~ Pablo Soler${}^{3}$}}
\vskip 0.8cm
{\it {${}^1$ Princeton University}}\\
{\it {Princeton, NJ 08544, USA}}\\
mkiermai@princeton.edu
\vskip .3cm
{\it {${}^2$ Institute of Physics, University of Tokyo}}\\
{\it {Komaba, Meguro-ku, Tokyo 153-8902, Japan}}\\
okawa@hep1.c.u-tokyo.ac.jp
\vskip .3cm
{\it {${}^3$ Instituto de F\'{i}sica Te\'{o}rica UAM/CSIC}}\\
{\it {Universidad Aut\'{o}noma de Madrid C-XVI }}\\
{\it {Cantoblanco, 28049 Madrid, Spain}}\\
pablo.soler@uam.es
\vskip 0.8cm
{\bf Abstract}
\end{center}
\noindent
We construct  analytic solutions of open string field theory
using boundary condition changing (bcc) operators.
We focus on bcc operators
with vanishing conformal weight
such as those for regular marginal deformations of the background.
For any Fock space state $\phi$, the component string field $\langle\phi, \Psi\rangle$
of the solution $\Psi$ exhibits a remarkable factorization property:
it is given by the matter three-point function of $\phi$ with
a pair of bcc operators,
multiplied by a universal function that
only depends on the conformal weight
of $\phi$.
This universal function is given by a simple integral expression
that can be computed once and for all.
The three-point functions with bcc operators are thus the only needed physical input
 of the particular open string background described by the solution.
We illustrate our solution with the example of the rolling tachyon profile,
for which we prove convergence analytically.
The form of our solution,
which involves bcc operators instead of explicit insertions of the marginal operator,
can be
a natural starting point for the construction of analytic solutions
for arbitrary backgrounds.

\medskip

\end{titlepage}

\renewcommand{\baselinestretch}{1.05}
\tableofcontents

\section{Introduction and summary}
\setcounter{equation}{0}

In the perturbative world-sheet formulation of string theory,
consistent backgrounds are described
by conformal field theories in two dimensions.
In the nonperturbative
formulation of string theory
we are searching for,
we expect that the requirement of conformal invariance
in the world-sheet theory
is reproduced from the classical equation of motion
in the spacetime theory.
String field theory is a candidate
for such nonperturbative
formulations,
and we expect a correspondence between
the space of conformal field theories
and the space of classical solutions.

In the case of the open string, a consistent background
is given by a choice of boundary conformal field theory (BCFT),
and different open string backgrounds correspond
to different conformal boundary conditions.
The change of boundary conditions can be described
by insertions of boundary condition changing
 (bcc)
operators
in the original BCFT.
For example, the change of the boundary conditions
on a segment of the world-sheet boundary from a point~$a$ to a point~$b$
can be described by inserting a pair of bcc operators
$\sigma_L (a)$ and $\sigma_R (b)$.
While bcc operators are generically not local,
they transform as primary fields under conformal transformations.

A deformation of boundary conditions is called marginal when the new
BCFT  is continuously connected to the original
BCFT by a one-parameter family of conformal boundary conditions.
A primary field $V(t)$ of weight one in the matter sector generates
an infinitesimal deformation of the BCFT,
and conformal invariance is preserved to linear order
in the deformation parameter which we denote by $\lambda$.
When operator products of the marginal operator $V(t)$ are regular,
finite deformations also preserve conformal invariance
and thus the operator $V(t)$ generates a family
of boundary conditions parameterized by $\lambda$.
In this case, the change of boundary conditions on a segment $[ \, a,b \, ]$
can be implemented by\footnote
{
If operator products of the marginal operator $V(t)$ are singular,
the conformal invariance can be violated
at higher order in $\lambda$.
When finite deformations preserve conformal invariance,
the deformation is called exactly marginal.
In this case, the change of the boundary conditions
can be implemented by renormalizing the exponential operator
in~(\ref{expVab}) appropriately.
See~\cite{Kiermaier:2007vu} for explicit examples.
}
\begin{equation}\label{expVab}
\sigma_L(a) \, \sigma_R(b)
~=~\exp \, \biggl[ \, \lambda \int_a^bdt\,V(t) \, \biggr]\,.
\end{equation}
The bcc operators associated with such regular marginal deformations
have vanishing conformal weights and satisfy
\begin{equation}\label{OPEsLsR}
\lim_{\epsilon \to 0} \sigma_L(0) \, \sigma_R(\epsilon) = 1 \,, \qquad
\lim_{\epsilon \to 0} \sigma_L(a) \, \sigma_R(b) \,
\sigma_L(b+\epsilon) \, \sigma_R(c)
= \sigma_L(a) \, \sigma_R(c) \,.
\end{equation}

\smallskip

The correspondence between conformal invariance in the world-sheet theory
and the equation of motion in the spacetime theory can therefore be restated
in the case of the open string
as a correspondence between a pair of bcc operators
and a solution to open string field theory (OSFT).
The equation of motion for open bosonic string field theory~\cite{Witten:1985cc}
is given by
\begin{equation}\label{EOM}
    Q\Psi+\Psi \ast \Psi \,=\,0\,,
\end{equation}
where $\Psi$ is an open string field of ghost number one,
$Q$ is the BRST operator,
and the symbol $\ast$ denotes multiplication of string fields
using Witten's star product.
 Since the construction of an analytic solution to~(\ref{EOM}) by Schnabl~\cite{Schnabl:2005gv},
an impressive amount of analytic results for string field theory have been
obtained~\cite{Okawa:2006vm, Fuchs:2006hw, Fuchs:2006an, Rastelli:2006ap, Ellwood:2006ba,Fuji:2006me,
Fuchs:2006gs, Okawa:2006sn, Erler:2006hw, Erler:2006ww, Schnabl:2007az, Kiermaier:2007ba, Erler:2007rh,
Okawa:2007ri, Fuchs:2007yy, Okawa:2007it, Ellwood:2007xr,Kishimoto:2007bb, Fuchs:2007gw,
Kiermaier:2007vu, Erler:2007xt, Rastelli:2007gg, Kiermaier:2007ki, Kwon:2007mh, Takahashi:2007du,Kiermaier:2007jg,Kwon:2008ap,Hellerman:2008wp,Ellwood:2008jh,Kawano:2008ry, Aref'eva:2008ad, Ishida:2008jc,Kawano:2008jv,Kiermaier:2008jy,Fuchs:2008cc,Asano:2008iu,Kishimoto:2008zj,Kiermaier:2008qu,Barnaby:2008pt, Aref'eva:2009ac, Kishimoto:2009cz,Ellwood:2009zf,Aref'eva:2009sj, Arroyo:2009ec,
Kroyter:2009bg,Erler:2009uj,AldoArroyo:2009hf, Beaujean:2009rb, Arroyo:2010fq, Zeze:2010jv, Schnabl:2010tb, Aref'eva:2010yd, Zeze:2010sr, Arroyo:2010sy, Erler:2010pr, Bonora:2010hi}.
These results partially illuminate the connection between the BCFT and OSFT descriptions of open string backgrounds. In particular, inspired by Ellwood's interpretation~\cite{Ellwood:2008jh}
of the gauge-invariant observable~\cite{Hashimoto:2001sm,Gaiotto:2001ji}
as the closed-string tadpole, an OSFT construction of the BCFT boundary state associated with known analytic solutions was presented
in~\cite{Kiermaier:2008qu}.

The other direction of the correspondence, namely, the construction of OSFT solutions associated with a given BCFT remains illusive.
Ideally, we would like to find a systematic construction of an OSFT solution from any given pair of bcc operators $\sigma_L$ and $\sigma_R$. Partial progress in this direction was achieved in~\cite{Kiermaier:2007vu},
where analytic solutions for general marginal deformations were constructed from
the bcc operators~(\ref{expVab}).
Unfortunately, the solution in~\cite{Kiermaier:2007vu} does not seem to be
the most promising starting point to construct analytic solutions for more general open string backgrounds. First of all, it was crucial for the construction in~\cite{Kiermaier:2007vu} to expand
the nonlocal operator
$\sigma_L(a)\sigma_R(b)$ in powers of the deformation parameter $\lambda$. For bcc operators that
describe
 generic open string backgrounds,
no such expansion
 parameter is
available and there are no straightforward ways to generalize the construction.
The second problem is more technical in nature.
The solution in~\cite{Kiermaier:2007vu} is constructed from wedge-based states\footnote
{We denote wedge states~\cite{Rastelli:2000iu} with operator insertions by wedge-based states.}
of integer width. In particular, the shortest wedge-based state appearing in the solution has
nonvanishing width.
As the  wedge width is additive under star multiplication, the operators inserted on this
 shortest
wedge state must be BRST-closed to satisfy the equation of motion~(\ref{EOM}). For generic open string backgrounds, however, there are no natural candidates for such operator insertions.\footnote{It is sometimes possible to construct time-dependent solutions from a {\em relevant} operator $\tilde{V}$ that triggers
a flow to a different background
by dressing the relevant operator with  $e^{\omega X^0}$ where $\omega$ is chosen to make $\tilde{V}e^{\omega X^0}$
exactly marginal~\cite{Bagchi:2008et}.
However, extracting the final state of this flow from
the late-time asymptotics
of such solutions is nontrivial.
}

There is another unsatisfactory feature shared by all known analytic solutions for marginal deformations.
When we calculate a coefficient
of the solution $\Psi$ given by the BPZ inner product
$\<\phi, \Psi\>$ for a state $\phi$ in the Fock space, one needs explicit knowledge of all $n$-point matter correlation functions
\begin{equation}
    \Bigl\langle\phi_m(0)V(t_1)V(t_2) \ldots
    V(t_n)\Bigr\rangle_{\rm UHP,\, matter}\,,
\end{equation}
where $\phi_m$ is the matter part of $\phi$ and UHP stands for upper half-plane.
These correlators are necessary for $\<\phi, \Psi\>$ at order $\lambda^n$ and are integrated over $t_i$ in a particular way. Therefore  the coefficient $\<\phi, \Psi\>$ has to be calculated from scratch for each choice of the matter operator  $\phi_m$ and the marginal operator $V$.
All information about a change in boundary conditions by $\sigma_L(a)\sigma_R(b)$, however, should in principle be captured entirely by the matter three-point functions\footnote{
Operator products of the bcc operators with other operators on the boundary
generate different operator insertions at the points
where the boundary conditions are changed.
We may need the information on these operator insertions
for more general solutions than regular marginal deformations.
}
\begin{equation}\label{Cphi}
    C_\phi=\bigl\langle\phi_m(0)\sigma_L(1)\sigma_R(\infty)\bigr\rangle_{\rm UHP,\, matter}\,\,.
\end{equation}
This aspect is obscured in all previously known analytic solutions for marginal deformations.

\bigskip

In this paper, we present solutions for regular marginal deformations
without any of the unsatisfactory features mentioned above.
The solution consists of wedge-based states,
and its operator insertions depend on the matter sector
only through bcc operators and their BRST transformations.
The solution is given by
\begin{equation}
\begin{split}\label{Psi2}
\Psi &=-\frac{1}{\sqrt{1-K}} \, (Q\sigma_L) \,
\frac{1}{1-K}\sigma_R \,(1-K)
Bc\frac{1}{\sqrt{1-K}} \,,
\end{split}
\end{equation}
where $K$, $B$, $c$, $\sigma_L$, and $\sigma_R$ are states based on the wedge state of zero width
with a line integral of the energy-momentum tensor and the $b$ ghost
for $K$ and $B$, respectively,
and with a local insertion of the $c$ ghost, $\sigma_{L}(t)$, and $\sigma_R(t)$
for $c$, $\sigma_L$, and $\sigma_R$,
respectively.\footnote{
 A precise definition of $K$, $B$, and $c$ is given below around~(\ref{Kdef}).  We follow the conventions of~\cite{Okawa:2006vm},
but the states
are rescaled as
 $K_{\rm here} = (\pi/2) \, K_{\rm there}$,
 $B_{\rm here} = (\pi/2) \, B_{\rm there}$,
 and
 $c_{\rm here} = (2/\pi) \, c_{\rm there}$.}
The state $K$ is the BRST transformation of $B$,
and the wedge state
$W_\alpha$ of width $\alpha$ is generated from $K$ as $W_\alpha = e^{\alpha K}$.

The solution $\Psi$ in~(\ref{Psi2})
is a special case of a class of analytic solutions for regular marginal deformations constructed by Erler~\cite{Erler:2007rh}, just as the ``simple'' analytic solution for tachyon condensation of~\cite{Erler:2009uj} is a special case
of a class of solutions in~\cite{Okawa:2006vm}.
It is interesting to note that exactly the same replacement $e^{K}\! \to 1/(1-K)$,
which was used in~\cite{Erler:2009uj} to transform the Schnabl-gauge solution,
also appears in our analysis.
Up to scaling of $K$,
it is the unique replacement that gives a solution based on bcc operators.\footnote
{
If we allow infinitely many bcc operators, there might be more solutions.
We would like to thank Ted Erler for discussion on this point.
It might be interesting to explore such possibilities
when we consider generalization to bcc operators with singular operator products.
}
Using Laplace transforms
of $1/(1-K)$ and $1/\sqrt{1-K}$\,, we can express the solution
 (\ref{Psi2})
in terms of wedge-based states:
\begin{equation}
\Psi
= -\int_0^\infty dr \int_0^\infty ds \int_0^\infty dt \, \frac{e^{-r-s-t}}{\pi\sqrt{rt}} \
e^{rK} ( Q\sigma_L ) \, e^{sK} \sigma_R \, (1-K)Bc \, e^{tK} \,.
\end{equation}
Note that no expansion of the bcc operators in $\lambda$ is necessary to define the solution. Furthermore, the solution takes the form of an integral over wedge-based states of width in the entire range
$[ \, 0,\infty)$\,,
and we thus expect it to be a natural starting point
for the construction of analytic
solutions for more general backgrounds.\footnote{
See~\cite{Ellwood:2009zf,Bonora:2010hi} for other interesting approaches to this problem.
}

\medskip

The calculation of coefficients $\<\phi,\Psi\>$ for this solution reduces to a simple evaluation of
the three-point function $C_\phi$ in~(\ref{Cphi}).
For example,
consider any operator $\phi$ of the form
\begin{equation}
    \phi= -c\partial c \, \phi_m\,,
\end{equation}
where $\phi_m$ is a matter primary field of weight $h\geq 1$.
The coefficient $\<\phi,\Psi\>$ in this case is simply given by
\begin{equation}\label{universal}
    \bigl\langle\, \phi\,,\,\Psi\,\bigr\rangle~=~ C_\phi\,g(h)\,,
\end{equation}
where $g(h)$ is a universal function of the weight $h$ of $\phi_m$,
but otherwise it is
independent of the particular choice of $\phi_m$  or the marginal operator $V$.\footnote{
 The generalization
to operators $\phi$ with different ghost sectors and matter descendant fields is straightforward, and different universal functions of $h$ can be obtained in this case.}
The explicit form of the function $g(h)$ is given by
\begin{equation}
\begin{split}
g(h)~&=~ \frac{h(h-1)}{2\pi}\,
\int_{\frac{1}{2}}^{\infty} dx \int_0^\infty ds \int_{\frac{1}{2}}^\infty dy \,
\frac{e^{1-x-s-y}}
{\sin^2 \theta_{s} \sqrt{( x-\tfrac{1}{2}) ( y-\tfrac{1}{2})}} \,
\biggl| \, \frac{2\sin \theta_{s}}
{L\sin \theta_{x} \sin \theta_{y}} \, \biggr|^h \\
& \hskip4.8cm \times
\Bigl [ \, \theta_{y} \,
\sin^2 \theta_{x}
+ \theta_{x} \,
\sin^2 \theta_{y}
- \sin{\theta_{x}} \, \sin{\theta_{s}}\, \sin{\theta_{y}} \,
\Bigr]
\qquad\text{ for }~ h>1\,,
\end{split}
\end{equation}
where $L=x+s+y$, and
 $\theta$ with a subscript is defined by
$\theta_\ell=\frac{\ell}{L}\,\pi$.
At $h=1$ and for large $h$,
the function $g(h)$
takes the form
\begin{equation}
    g(1)\,=\,1\,, \qquad~~ g(h)\,\sim\,
    \biggl( \frac{8}{\pi} \biggr)^h
    ~~
    \text{ for }
    ~
    h\gg1\,.
\end{equation}
This exact result for the large-$h$ behavior of $g(h)$ will allow us to prove the convergence of the tachyon profile of the rolling tachyon solution.
A plot of $ g(h)$ is
presented in figure~\ref{gplot} of section~\ref{secuniversal}.

This paper is organized as follows. In section~\ref{secderivation} we derive the solution~(\ref{Psi2}) as a special case of a class of analytic solutions for regular marginal deformations constructed by Erler~\cite{Erler:2007rh}.
In section~\ref{secuniversal} we establish the universal behavior~(\ref{universal}) of coefficients $\<\phi,\Psi\>$ and study
the asymptotic behavior of the function $g(h)$.
As an application, the results for $g(h)$ are then used in section~\ref{sectachyon} to extract the tachyon profile from the rolling tachyon solution.

\section{Derivation of the solution}
\setcounter{equation}{0}\label{secderivation}

Analytic solutions for marginal deformations were first constructed
in~\cite{Schnabl:2007az,Kiermaier:2007ba}
when operator products of the marginal operator $V(t)$
are regular.
The solutions are given
as a perturbative expansion in the deformation parameter $\lambda$:
\begin{equation}\label{Psi}
    \Psi=\sum_{n=1}^\infty \lambda^n \Psi^{(n)}\,.
\end{equation}
Expressed as a conformal field theory (CFT) correlator,
$\Psi^{(n)}$
in Schnabl gauge~\cite{Schnabl:2007az,Kiermaier:2007ba} is given by
\begin{equation}
\label{KORZ-solution}
\begin{split}
\langle \, \phi, \Psi^{(n)} \, \rangle
&  = \int_0^1 \hskip-3pt dt_1 \int_0^1 \hskip-3pt dt_2 \ldots
\int_0^1 \hskip-3pt dt_{n-1} \,
\langle \, f \!\circ\! \phi (0) \, c V (1) \, {\cal B} \,
c V (1+t_1) \, {\cal B} \, c V (1+t_1+t_2) \, \ldots \\
& \qquad \qquad \qquad \qquad \qquad\qquad\, {}\times
{\cal B} \, c V (1+t_1+t_2+ \ldots +t_{n-1}) \,
\rangle_{{\cal W}_{1+t_1+t_2+ \ldots +t_{n-1}}}\, .
\end{split}
\end{equation}
Here and in what follows we denote a generic state
in the Fock space by $\phi$ and its corresponding operator
in the state-operator mapping by
 $\phi (\xi)$.
We denote
the conformal transformation of $\phi (\xi)$ under the map $f(\xi)$
by $f \circ \phi (\xi)$, where
\begin{equation}
\label{arctan}
f(\xi)= \frac{2}{\pi} \, \arctan \xi \,.
\end{equation}
The correlation function is evaluated
on the wedge surface ${\cal W}_\alpha$
with $\alpha \ge 0$,
which is
the semi-infinite strip
on the upper half-plane
of~$z$
between the vertical lines
$\Re(z)=-\frac{1}{2}$ and $\Re(z)=\frac{1}{2}+\alpha$
with these lines identified by translation.
The operator ${\cal B}$ is a line integral of
the $b$ ghost
defined by
\begin{equation}\label{defBCFT}
    {\cal B}=
\int_{i \infty}^{-i\infty}
\frac{dz}{2\pi i}\,b(z) \,,
\end{equation}
where we used the doubling trick.
Its BRST transformation is given by
\begin{equation}\label{defKCFT}
{\cal K} =
\int_{i \infty}^{-i\infty} \frac{dz}{2\pi i } \,
\oint \frac{dw}{2\pi i } \, j_B (w) \, b(z)
= \int_{i \infty}^{-i\infty}
\frac{dz}{2\pi i}\,T(z) \,,
\end{equation}
where $j_B$ is the BRST current, $T$ is the energy-momentum tensor,
and the contour of the integral over $w$ encircles $z$ counterclockwise.
The line integral ${\cal K}$ of
the energy-momentum tensor
is the generator of infinitesimal changes in the width of
the wedge state $W_\alpha$ defined by
\begin{equation}
\langle \, \phi, W_\alpha \, \rangle
= \langle \, f \!\circ\! \phi (0)
\rangle_{{\cal W}_{\alpha}}\, .
\end{equation}
Indeed, we have
\begin{equation}\label{CFTKgenW}
\langle \, \phi, \del_\alpha W_\alpha \, \rangle
 = \langle \, f \!\circ\! \phi (0)\, {\cal K}\,
\rangle_{{\cal W}_{\alpha}}\, .
\end{equation}
\medskip

The solution~(\ref{KORZ-solution}) can also be expressed in an algebraic language
without referring to explicit CFT correlators. We denote by $K$ the string field that generates
the wedge states
through the relation
\begin{equation}
    W_\alpha=e^{\alpha K}\,.
\end{equation}
We can think of $K$
as a wedge state of zero width with an insertion of ${\cal K}$:
\begin{equation}\label{Kdef}
 \langle \, \phi, K \, \rangle
 =
 \langle \, f \!\circ\! \phi (0)\,{\cal K}\,
 \rangle_{{\cal W}_{0}}\, .
\end{equation}
Note that the identity~(\ref{CFTKgenW}) is manifest in this algebraic language:
\begin{equation}
    \del_\alpha e^{\alpha K}= e^{\alpha K}K\,.
\end{equation}
We denote
analogous wedge-based states of zero width with insertions of
${\cal B}$,
 $c(\frac{1}{2})$, and $V(\frac{1}{2})$ on the wedge surface ${\cal W}_0$ by $B$, $c$, and $V$, respectively.
These states satisfy the following relations:
\begin{equation}
\begin{split}\label{KBcAlgebra}
[K,B] = 0\,,\qquad [B,V]=0\,,\qquad [c,V]=0\,,\qquad c^2=0\,,\qquad B^2=0\,,\qquad\{B,c\} = 1
 \,.
\end{split}
\end{equation}
The BRST transformation $Q$ acts on these states in the following way:
\begin{equation}\label{BRSTtransformations}
Q B = K \,, \qquad Q K = 0 \,,\qquad Q c = cKc \,, \qquad Q V = [K,cV] \,.
\end{equation}
It follows that
$Q \, ( cV ) =0$,
which expresses the marginality of the operator $V$.

In this algebraic language, the solution~(\ref{KORZ-solution}) takes the form
\begin{equation}
\Psi^{(n)}=e^{K/2} cV\Bigl(\int_0^1dt\,B e^{t K}cV\Bigr)^{n-1} e^{K/2}=e^{K/2} cV\Bigl(B\frac{e^{K}-1}{K}cV\Bigr)^{n-1} e^{K/2}\,.
\end{equation}
The perturbative series in $\lambda$ that defines $\Psi$ in~(\ref{Psi}) can then be summed to obtain
\begin{equation}\label{PsiKORZ}
    \Psi=\lambda e^{K/2} cV\Bigl(1-\lambda B\frac{e^{K}-1}{K}cV\Bigr)^{-1} e^{K/2}\,.
\end{equation}
 In~\cite{Erler:2007rh}, Erler showed that
 gauge-equivalent
 solutions can be obtained if one replaces $e^{K/2}$ appearing in~(\ref{PsiKORZ}) by a general function $f(K)$:
\begin{equation}\label{erler}
\Psi = f(K) \, \lambda \, cV \,
\biggl[ \, 1 - B \, \frac{f(K)^2-1}{K} \,
\lambda \, cV \, \biggr]^{-1} f(K) \,.
\end{equation}
To avoid the wedge state $W_\alpha = e^{\alpha K}$
with negative $\alpha$, we require $f(K)$ to take the following form:
\begin{equation}
f(K) = \int_0^\infty dt \, \tilde{f}(t) \, e^{tK} \,.
\end{equation}
We are looking for
a choice of $f(K)$
 that
allows us to express $\Psi$ in terms of bcc operators.
Consider the wedge state with a boundary condition
modified by a marginal deformation generated by the operator $V(t)$.
An insertion of
\begin{equation}\label{modwedgeoperator}
\exp \, \biggl[ \, \lambda \int_a^b dt \, V(t) \, \biggr]
= 1 + \lambda \int_a^b dt_1 \, V(t_1)
+ \lambda ^2 \int_a^b dt_1 \int_{t_1}^b dt_2 \, V(t_1) \, V(t_2) \,
+ \ldots
\end{equation}
corresponds to
changing the wedge state $e^{\alpha K}$ with $\alpha = b-a$
to $U_\alpha$ given by
\begin{equation}
\label{modwedgestate}
U_\alpha =  e^{\alpha K}
+ \lambda \int_0^\alpha d t_1 \,
e^{ t_1 K} \, V \, e^{(\alpha -t_1) K}
+ \lambda^2 \int_0^\alpha d t_1 \int_{t_1}^\alpha d t_2 \,
e^{ t_1 K} \, V \,
e^{(t_2-t_1) K} \, V \,
e^{(\alpha -t_2) K} + \ldots \,.
\end{equation}
When $\alpha$ is small, $U_\alpha$ reduces to
\begin{equation}
U_\alpha = 1+ \alpha \, ( K + \lambda V ) + {\cal O}(\alpha^2) \,.
\end{equation}
 We can also show that $U_{\alpha+\beta}$ factorizes as
 \begin{equation}
 U_{\alpha+\beta} ~=~  U_\alpha \, U_\beta \,,
 \end{equation}
 which is obvious from the structure of $U_\alpha$
 similar to that of the path-ordered exponential.
 {}From these two properties, we conclude that
\begin{equation}
U_\alpha = e^{\alpha ( K + \lambda \, V )} \,.
\end{equation}
Therefore, the wedge state with the modified boundary condition
is given by $e^{\alpha ( K + \lambda \, V )}$.
In the language of bcc operators,
this can be stated as follows:
\begin{equation}\label{sigmas}
 \ldots \, e^{\alpha ( K + \lambda \, V )} \,
 \ldots
 ~=~ \ldots\, \sigma_L \, e^{\alpha K} \, \sigma_R \,
 \ldots \,,
\end{equation}
where the dots $\ldots$ represent
arbitrary wedge-based states
with the boundary conditions of the undeformed BCFT.
Let us next consider the BRST transformation of the state $e^{\alpha ( K + \lambda \, V )}$.
We use the formula
\begin{equation}
\delta \, e^{\alpha M} = \int_0^\alpha dt \, e^{t M} \, \delta M \, e^{(\alpha-t) M}
\end{equation}
for any derivation
$\delta (M_1 M_2) = (\delta M_1) \, M_2 + M_1 \, (\delta M_2)$
with respect to the multiplication under consideration.
For star products of Grassmann-even states, the BRST transformation
$Q (M_1 M_2) = (Q M_1) \, M_2 + M_1 \, (Q M_2)$
and the commutator
$[ \, N,\,  M_1 M_2 \, ] = [ \, N,\, M_1 \, ] \, M_2 + M_1 \, [ \, N,\, M_2 \, ]$
are such derivations.
Since
\begin{equation}
Q \, (K+\lambda V) = [ \, K+\lambda V,\,  \lambda \, c V \, ] \,,
\end{equation}
which follows from~(\ref{BRSTtransformations}), we find that
\begin{equation}
\begin{split}
Q \, e^{\alpha ( K + \lambda \, V )}
& = \int_0^\alpha dt \, e^{t ( K + \lambda \, V )} \,
Q \, (K+\lambda V) \, e^{(\alpha-t) ( K + \lambda \, V )} \\
& = \int_0^\alpha dt \, e^{t ( K + \lambda \, V )} \,
[ \, K+\lambda V,\,  \lambda \, c V \, ] \, e^{(\alpha-t) ( K + \lambda \, V )} \\[.5ex]
& = e^{\alpha ( K + \lambda \, V )} \, \lambda \, c V
-\lambda \, c V \, e^{\alpha ( K + \lambda \, V )} \,.
\end{split}
\end{equation}
In the language of bcc operators,
we can write
\begin{equation}
\begin{split}\label{Qsigmas}
\ldots \, e^{\alpha ( K + \lambda \, V )}( \lambda \, cV ) \, \ldots
~&=~\ldots \, \sigma_L \, e^{\alpha K} \, ( Q \, \sigma_R ) \, \ldots\,,\\
 \ldots \, (-\lambda \, cV ) \, e^{\alpha ( K + \lambda \, V )} \, \ldots~&=~
\ldots \, ( Q \, \sigma_L ) \,
e^{\alpha K} \, \sigma_R \, \ldots \,.
\end{split}
\end{equation}

Our goal is
 to find a choice of $f(K)$ such that
the solution $\Psi$ can be written
in terms of $\sigma_L$ and $\sigma_R$
 and their BRST transformations
without using $V$ explicitly.
This is achieved if $\lambda V$ only appears
in the solution through the combination
$h(K+\lambda V)$, $h(K+\lambda V)(\lambda cV)$, or $(\lambda cV)h(K+\lambda V)$
with arbitrary functions $h(x)$ in the following form:
  \begin{equation}\label{gLaplace}
    h(x)=\int_0^\infty d\alpha\, \tilde h(\alpha) \, e^{\alpha x}\,.
  \end{equation}
This ensures that $h(K+\lambda V)$ has support on wedge states of
 nonnegative width.
 Indeed,
 \begin{equation}
        h(K+\lambda V)=\int_0^\infty d\alpha \, \tilde h(\alpha) \, e^{\alpha (K+\lambda V)}\,.
 \end{equation}
 It then follows from~(\ref{sigmas}) and~(\ref{Qsigmas}) that
 \begin{equation}
  \boxed{
 \begin{aligned}
 \phantom{\Bigl(}
 \ldots \, h( K + \lambda \, V ) \, \ldots ~&=~ \ldots\, \sigma_L \, h( K) \, \sigma_R \, \ldots \,,\\
\ldots \, h( K + \lambda \, V )( \lambda \, cV ) \, \ldots ~&=~\ldots \, \sigma_L \, h( K) \, ( Q \, \sigma_R ) \, \ldots\,,\\
 ~~
 \ldots \, (\!{}-\! \lambda \, cV ) \, h( K + \lambda \, V ) \, \ldots~&=~
\ldots \, ( Q \, \sigma_L ) \,
h(K) \, \sigma_R \, \ldots \,.
 \phantom{\Bigl(}~
\end{aligned}
  }
\end{equation}

 To find a choice of
 $f(K)$ that brings the solution~(\ref{erler}) into this form, it is convenient to
 first transform $\Psi$ slightly.
 As shown in
 appendix~\ref{secaltderiv},
 the solution can be written as
\begin{equation}
\begin{split}\label{alternative}
\Psi & = f(K) \, \lambda \, cV \,
\biggl[ \, 1 - \frac{f(K)^2-1}{K} \,
\lambda \,V \, \biggr]^{-1} Bc \, f(K) \,.
\end{split}
\end{equation}
To obtain an expression for $\Psi$ in terms of
 (a finite number of)
bcc operators, we choose
\begin{equation}
f(K) = \frac{1}{\sqrt{1-K}} \,.
\end{equation}
The derivation of this is presented in appendix~\ref{f^2=1/(1-K)}.
With this choice, we have
\begin{equation}\label{newsolution}
\boxed{
\begin{aligned}
 \phantom{\Biggl(}
\Psi
&=~ \frac{1}{\sqrt{1-K}} \, \lambda \, cV \,
\frac{1}{1-K-\lambda \, V} (1-K)
Bc\frac{1}{\sqrt{1-K}}\\[-1ex]
&=-\frac{1}{\sqrt{1-K}} \, (Q\sigma_L) \,
\frac{1}{1-K}\sigma_R \,(1-K)
Bc\frac{1}{\sqrt{1-K}} \,,
 \phantom{\Biggl(}
\end{aligned}
~}
\end{equation}
where we used the identity
\begin{equation}
\biggl[ \, 1 - \frac{1}{1-K} \, \lambda \, V \, \biggr]^{-1}
= \frac{1}{1-K-\lambda \, V} (1-K) \,.
\end{equation}

It is easy to expand this solution as a superposition of
wedge-based states.
Using
\begin{equation}\label{dressed}
    \frac{1}{1-K}=\int_0^\infty ds\, e^{-s}e^{sK}\,,  \qquad
    \frac{1}{\sqrt{1-K}}=\int_0^\infty ds\,  \frac{e^{-s}}{\sqrt{\pi s}} \,
    e^{sK}\,,
\end{equation}
we obtain
\begin{equation}
\Psi
=-\int_0^\infty dr\int_0^\infty ds\int_0^\infty dt \frac{e^{-r-s-t}}{\pi\sqrt{rt}}\,e^{rK} \, (Q\sigma_L) \,
e^{sK}\sigma_R \,(1-K)
Bc\,e^{tK}\,.
\end{equation}
Let us also present the solution $\Psi$
in the CFT language.
 Recalling that the line integrals associated with $K$ and $B$ are denoted by
 ${\cal K}$
and ${\cal B}$, respectively, we have
\begin{equation}\label{solBCCO}
        \bigl\langle\phi\,,\,\Psi\bigr\rangle=
        -\!\!\int_0^\infty\!\!\!\!\!dr\!\int_0^\infty\!\!\!\!\!\! ds\!\int_0^\infty\!\!\!\!\!\! dt\, \frac{e^{-r-s-t}}{\pi\sqrt{rt}}
        \Bigl\langle f\circ \phi(0)~Q\sigma_L\bigl(\tfrac{1}{2}+r\bigr) \,\sigma_R\bigl(\tfrac{1}{2}+r+s\bigr)\,
        (1-{\cal K})\,{\cal B}\,
        c\bigl(\tfrac{1}{2}+r+s\bigr)\!\Bigr\rangle_{{\cal W}_{r+s+t}}\!\!\!\!.\,\,\,
\end{equation}

The solution $\Psi$
 satisfies the reality condition on the string field~\cite{Gaberdiel:1997ia}, but
it is not manifest in~(\ref{newsolution}).
This can be seen as follows:
\begin{equation}
\begin{split}
\Psi
&=~ \frac{1}{\sqrt{1-K}} \, \lambda \, cV \,
\frac{1}{1-K-\lambda \, V} \,
\bigl[ \, (1-K -\lambda \, V ) +\lambda \, V \, \bigr] \,
Bc \, \frac{1}{\sqrt{1-K}} \\
&=~ \frac{1}{\sqrt{1-K}} \, \lambda \, cV \,
\frac{1}{\sqrt{1-K}}
+ \frac{1}{\sqrt{1-K}} \, \lambda \, cV \,
\frac{B}{1-K-\lambda \, V} \,
\lambda \, cV \, \frac{1}{\sqrt{1-K}} \,.
\end{split}
\end{equation}
This form is manifestly symmetric when we reverse the order of multiplication of string fields and thus satisfies the reality condition of~\cite{Gaberdiel:1997ia},
 which
guarantees that the string field theory action is real.
In terms of bcc operators, this can be written in the following form:
\begin{equation}\label{symmetric-form}
\boxed{~~
\Psi
\,=\,-\frac{1}{\sqrt{1-K}} \, ( Q \sigma_L)\, \sigma_R \frac{1}{\sqrt{1-K}}
\,\,-\,\,\frac{1}{\sqrt{1-K}} \, ( Q \sigma_L) \,
\frac{B}{1-K}\,( Q \sigma_R)\,
\frac{1}{\sqrt{1-K}}\,.
~~}
\end{equation}
Although we arrived at this form from an expression that contained the marginal parameter $\lambda$ and operator $V$ explicitly,
 it is now written only in terms of $K$, $B$, $\sigma_L$, and $\sigma_R$.
 This is a solution to the equation of motion for any choice of bcc operators $\sigma_L$, $\sigma_R$ in the matter sector
 that satisfy the operator products~(\ref{OPEsLsR}), with $K$ and $B$ defined around~(\ref{Kdef}).

More generally,~(\ref{symmetric-form}) satisfies the equation of motion for any choice of three states $B$, $\sigma_L$, and $\sigma_R$  satisfying the relations
\begin{equation}
\label{KBsigma}
B^2 = 0 \,, \qquad
[ \, B, \sigma_L \, ] = 0 \,, \qquad
[ \, B, \sigma_R \, ] = 0 \,, \qquad
\sigma_L \, \sigma_R = 1 \,, \qquad
\sigma_R \, \sigma_L = 1 \,,
\end{equation}
and
$K=QB$ serving as a definition of $K$.
By considering the BRST transformation of each of
the relations in~(\ref{KBsigma}),
we find
\begin{equation}
\label{KBsigma-BRST}
\begin{split}
& [ \, K, B \, ] = 0 \,, \quad
\{ \, B, Q \sigma_L \, \} = [ \, K, \sigma_L \, ] = {}-[ \, 1-K, \sigma_L \, ] \,, \quad
\{ \, B, Q \sigma_R \, \} = [ \, K, \sigma_R \, ] = {}-[ \, 1-K, \sigma_R \, ] \,, \\[1ex]
& ( Q \sigma_L ) \, \sigma_R + \sigma_L \, ( Q \sigma_R ) = 0 \,, \qquad
( Q \sigma_R ) \, \sigma_L + \sigma_R \, ( Q \sigma_L ) = 0 \,.
\end{split}
\end{equation}
We can verify that the solution~(\ref{symmetric-form}) satisfies the equation of motion
{\it only} from (\ref{KBsigma}) and (\ref{KBsigma-BRST}),
and no explicit reference to $\lambda$ and $V$
or to the surface state definitions of $K$ and $B$ is necessary.
Incidentally, we do not need to assume that $( Q \sigma_L ) \, ( Q \sigma_R)$ vanishes,
while it does for regular marginal deformations we started with.

\section{Universal coefficients}
\setcounter{equation}{0}\label{secuniversal}

In the CFT language, the solution $\Psi$ is specified
by giving $\<\phi, \Psi\>$ for an arbitrary state $\phi$ in the Fock space.
We can choose a basis of states in the Fock space
such that the matter and ghost sectors are factorized.
In this section we demonstrate that,
when $\phi$ is in the factorized basis and its matter part is a primary field,\footnote{
It is straightforward
to generalize our analysis to descendant fields, which would result in
different
universal factors.
}~
the inner product $\<\phi, \Psi\>$
is given by a product of a universal $V$-independent factor
and a simple three-point function of the matter part of $\phi$
with the bcc operators $\sigma_L$ and $\sigma_R$.

Since we are considering bcc operators with vanishing conformal weight,
their BRST transformations are given by
$Q\sigma_L= c \partial\sigma_L$
and $Q\sigma_R= c \partial\sigma_R$.
The solution $\Psi$ in the form~(\ref{symmetric-form}) can then be written as
\begin{equation}\label{symPsi}
\Psi
\,=\,-\frac{1}{\sqrt{1-K}} \, (c \partial\sigma_L)\, \sigma_R \frac{1}{\sqrt{1-K}}
\,\,-\,\,\frac{1}{\sqrt{1-K}} \, (c \partial\sigma_L) \,
\frac{B}{1-K}\,(c\partial \sigma_R)\,
\frac{1}{\sqrt{1-K}}\,.
\end{equation}
Let us calculate the inner product $\<\phi, \Psi\>$ for $\phi = -c\partial c \, \phi_m$,
where $\phi_m$ is a matter primary field of weight $h$.
The first term in~(\ref{symPsi}) is a superposition of wedge states with a single insertion of $(c \partial\sigma_L)\, \sigma_R=-\lambda cV$, and its inner product with $\phi$ thus vanishes
unless $h=1$.
We postpone the special case $h=1$, and first consider the case $h>1$.
 It is then sufficient to evaluate the second term. We have
\begin{equation}\label{phiPsi}
\<\phi, \Psi\>\,=\,
-\int_{\frac{1}{2}}^{\infty} dx \int_0^\infty ds \int_{\frac{1}{2}}^\infty dy \,
\frac{e^{1-x-s-y}}
{\pi\sqrt{( x-\tfrac{1}{2}) ( y-\tfrac{1}{2})}}\,
\Bigl\langle \, f \!\circ\! \phi (0) \,
c \partial \sigma_L (x) \, {\cal B} \,
c \partial \sigma_R (x+s) \, \Bigr\rangle_{{\cal W}_{L-1}}\,,
\end{equation}
where we have defined
\begin{equation}
L = x+s+y \,.
\end{equation}
The correlator in~(\ref{phiPsi}) can be written in a matter-ghost factorized form as
\begin{equation}\label{factorized}
\begin{split}
 & \langle \, f \!\circ\! \phi (0) \, c \partial \sigma_L (x) \, {\cal B} \, c \partial \sigma_R (x+s) \, \rangle_{{\cal W}_{L-1}} \\
& = -\frac{\pi}{2} \,
\langle \, c \partial c (0) \,
c(x) \, {\cal B} \, c(x+s) \, \rangle_{{\cal W}_{L-1},\,{\rm ghost}} \,\times\,
\partial_a \partial_b \, \mla \, f \!\circ\! \phi_m (0) \,
\sigma_L (a) \, \sigma_R (b) \,
\mra_{{\cal W}_{L-1}}\biggr|_{a=x,\,b=x+s} \,,
\end{split}
\end{equation}
where we use $\mla \ldots
\mra$ to denote matter correlators $\langle \ldots
\rangle_{\rm matter}$.
Since $\phi_m$, $\sigma_L$, and $\sigma_R$ are
primary fields of weight $h$, $0$, and $0$, respectively,
we find\footnote{
Even when we write the matter and ghost sectors separately,
it should be understood that
we always perform conformal transformations for the combined system,
which has a vanishing central charge.
}
\begin{equation}
\label{h-dependence}
\begin{split}
\mla \, f \!\circ\! \phi_m (0) \,
\sigma_L (a) \, \sigma_R (b) \,
\mra_{{\cal W}_{L-1}}
&\, =\, \Bigl( \frac{2}{L} \Bigr)^h \mla \, f \!\circ\! \phi_m (0) \,
\sigma_L \Bigl( \frac{2 a}{L} \Bigr) \,
\sigma_R \Bigl( \frac{2 b}{L} \Bigr) \,
\mra_{{\cal W}_1} \\[.5ex]
& \,=\, \Bigl( \frac{2}{L} \Bigr)^h  \mla \, \phi_m (0) \,
\sigma_L \bigl( \tan \theta_a \bigr) \,
\sigma_R \bigl( \tan \theta_b \bigr) \,
\mra_{{\rm UHP}} \\[.5ex]
& \,=\, C_\phi \, \Bigl| \,\frac{2\,\sin \theta_{b-a}}
{L \sin \theta_a \sin \theta_b}  \Bigr|^h \,,
\end{split}
\end{equation}
where $\theta$ with a subscript is defined by
\begin{equation}
    \theta_\ell=\frac{\ell}{L}\,\pi\,,
\end{equation}
and
$C_\phi$ is a constant independent of $a$, $b$, and $L$.
It is related to
the coefficient of the matter three-point function of $\phi$, $\sigma_L$ and $\sigma_R$
as follows:
\begin{equation}
    \mla \, \phi_m (z_1) \,
\sigma_L (z_2) \,
\sigma_R ( z_3) \,
\mra_{{\rm UHP}}
~= ~C_\phi\,
\biggl|\frac{z_2-z_3}{(z_1-z_2)(z_1-z_3)}\biggr|^h\,.
\end{equation}
In other words, $C_\phi$ is
the matter three-point function with operators $\phi_m$, $\sigma_L$, and $\sigma_R$ inserted at $0$, $1$, and $\infty$, respectively:\footnote{
Since the weight of $\sigma_R$ vanishes,
we can simply send the position of $\sigma_R$ to infinity
without considering the conformal transformation $I(\xi) = -1/\xi$.
}
\begin{equation}
    C_\phi =  {}\mla \, \phi_m (0) \,
\sigma_L (1 ) \,
\sigma_R ( \infty) \,
\mra_{{\rm UHP}}\,.
\end{equation}
For the matter correlator in~(\ref{factorized}), we then have
\begin{equation}
\begin{split}\label{matter}
& \partial_a \partial_b \, \mla \, f \!\circ\! \phi_m (0) \,
\sigma_L (a) \, \sigma_R (b) \,
\mra_{{\cal W}_{L-1}}\biggr|_{a=x,\,b=x+s}
 \!=\, {}- C_\phi \,
\frac{\pi^2 h (h-1)}{L^2\sin^{2} \theta_{s}} \,
\biggl|  \, \frac{2\sin \theta_{s}}
{L\sin \theta_x \sin \theta_y} \, \biggr|^h\,.
\end{split}
\end{equation}
The ghost sector correlator takes the form
\begin{equation}
\begin{split}\label{ghost}
\langle \, c \partial c (0) \,
c(x) \, {\cal B} \, c(x+s) \, \rangle_{{\cal W}_{L-1},\,{\rm ghost}}
&\,=\,
-\frac{L^2}{\pi^3}\Bigl[ \, \theta_{y}\sin^2\theta_{x}+\theta_{x} \sin^2\theta_{y}
-\sin\theta_{x}\sin\theta_{s}\sin\theta_{y} \, \Bigr] \,.
\end{split}
\end{equation}
Combining the results for the matter and ghost correlators~(\ref{matter}) and~(\ref{ghost}),
we obtain
\begin{equation}
\boxed{\phantom{\Bigl(}
    \bigl\langle\phi\,,\,\Psi\bigr\rangle~=~C_\phi\, g(h) ~}
\end{equation}
with
\begin{equation}
\label{gfunction}
\begin{split}
g(h)~&=~ \frac{h(h-1)}{2\pi}\,
\int_{\frac{1}{2}}^{\infty} dx \int_0^\infty ds \int_{\frac{1}{2}}^\infty dy \,
\frac{e^{1-x-s-y}}
{\sin^2 \theta_{s} \sqrt{( x-\tfrac{1}{2}) ( y-\tfrac{1}{2})}} \,
\biggl| \,
\frac{2\sin \theta_{s}}
{L\sin \theta_{x} \sin \theta_{y}} \, \biggr|^h \\
& \hskip4.8cm
\times \Bigl [ \,
\theta_{y} \,
\sin^2 \theta_{x}
+ \theta_{x} \,
\sin^2 \theta_{y}
- \sin{\theta_{x}} \, \sin{\theta_{s}} \, \sin{\theta_{y}} \,
\Bigr]
\qquad\text{ for }~ h>1\,.
\end{split}
\end{equation}
This is the factorized form we mentioned before:
the inner product $\bigl\langle\phi\,,\Psi\bigr\rangle$
is given by a product of the three-point function $C_\phi$
and the universal function $g(h)$, which does not depend on $V$.

\begin{figure}[t]
\centering
\includegraphics[bb=0 0 10.59cm 6.84cm, width=.75\textwidth]{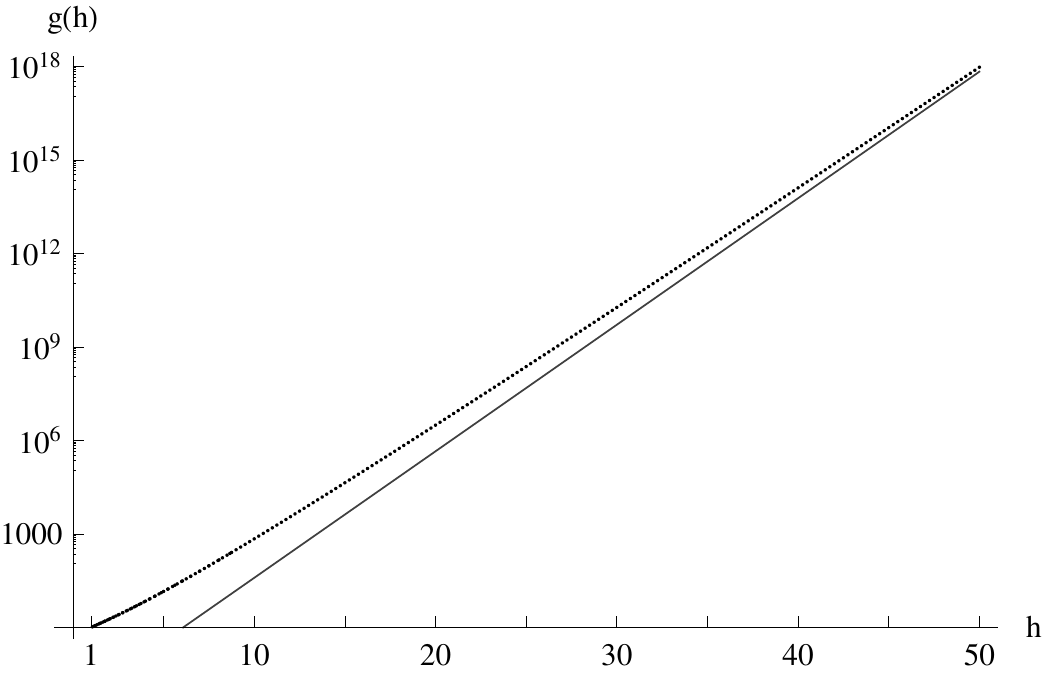}
\caption{\small
The black dots represent numerical evaluations of the function $g(h)$ given in eq.~(\ref{gfunction}). The gray line is the function $a(8/\pi)^h$, with the coefficient $a$ fitted
 to numerical evaluations of $g(h)$ in the range $1.25\leq h\leq 60$.
}
\label{gplot}
\end{figure}

When $h=1$,
the first term (and only the first term) of the solution in~(\ref{symPsi}) contributes
to $\<\phi,\Psi\>$.
It is given by
\begin{equation}
\<\phi, \Psi\>\,=\,
-\int_{\frac{1}{2}}^{\infty} dx \int_{\frac{1}{2}}^\infty dy \,
\frac{e^{1-x-y}}
{\pi\sqrt{( x-\tfrac{1}{2}) ( y-\tfrac{1}{2})}}\,
\Bigl\langle \, f \!\circ\! \phi (0) \,
c (\partial \sigma_L) \sigma_R (x) \, \Bigr\rangle_{{\cal W}_{x+y-1}}\,,
\end{equation}
and we find

\begin{equation}
    g(1)\,=1\,.
\end{equation}
The leading behavior of $g(h)$
at $h\gg 1$ can  be determined analytically.
We note that the dominant contribution to $g(h)$
in this limit comes
from the part of the integration region where
the factor $|\ldots|^h$
in the integrand is maximized. It is easy to see that
\begin{equation}
\frac{2\sin \theta_{s}} {L\sin \theta_{x} \sin \theta_{y}}  \,<\, \frac{8}{\pi}\,,
\end{equation}
with the bound saturated  at $x=y=\frac{1}{2}$ in the limit $s\to\infty$. We conclude that $g(h)$ behaves as
\begin{equation}\label{asyg}
    g(h)\,\sim\, \Bigl(\frac{8}{\pi}\Bigr)^h\qquad\text{ for }~~h\gg1\,.
\end{equation}
Figure~\ref{gplot} gives a plot of $g(h)$ on a logarithmic scale, together with the asymptote of its large-$h$ behavior.\footnote{
Numerical fits suggest that the function $\log[g(h)]$ takes the form $\log(8/\pi)h+{\cal O}(h^{1/3})$ at large $h$. It would be interesting to derive the curious subleading behavior $\sim h^{1/3}$ analytically.}

\section{Application to the rolling tachyon profile}
\setcounter{equation}{0}\label{sectachyon}

The rolling tachyon solution
represents
 the time-dependent process of
D-brane decay~[\citen{Sen:2002nu,Sen:2002in,Sen:2002vv,Moeller:2002vx,Larsen:2002wc,Lambert:2003zr,Fujita:2003ex,Gaiotto:2003rm,Erler:2004sy,Sen:2004nf,Coletti:2005zj},
\citen{Schnabl:2007az,Kiermaier:2007ba}].
It can be constructed by choosing $V(t)$ to be\footnote{
Here and in what follows
boundary normal ordering
for the exponential operator of $X^0$ is implicit.
}
\begin{equation}
V(t) = e^{\frac{1}{\sqrt{\alpha'}} X^0}\! (t) \,.
\end{equation}
For this choice, operator products $V(t_1) \, V(t_2) \, \ldots V(t_n)$
are regular~\cite{Schnabl:2007az,Kiermaier:2007ba}.
For example, the leading term of $V(t) \, V(0)$
in the limit $t \to 0$ is given by
\begin{equation}
V(t) \, V(0) \sim | \, t \, |^2 \, e^{\frac{2}{\sqrt{\alpha'}} X^0}\! (0) \,.
\end{equation}
We therefore have
\begin{equation}
\lim_{\epsilon \to 0} \, \exp \, \biggl[ \, \lambda \int_0^\epsilon dt \,V(t) \, \biggr]
= 1 \,,\quad
\lim_{\epsilon \to 0} \, \exp \, \biggl[ \, \lambda \int_a^b dt\,V(t) \, \biggr] \,
\exp \, \biggl[ \, \lambda \int_{b+\epsilon}^c dt\,V(t) \, \biggr]
=  \exp \, \biggl[ \, \lambda \int_a^c dt\,V(t) \, \biggr] \,,
\end{equation}
and thus the corresponding bcc operators satisfy
the two conditions in~(\ref{OPEsLsR}).\footnote
{
Another example of a marginal operator with regular operator products
is given by the lightcone-like operator $i \partial X^\pm$,
as mentioned in~\cite{Schnabl:2007az} and studied in~\cite{Kiermaier:2007ba}.
}
The magnitude of the deformation parameter $\lambda$ can be changed by time translation,
so all solutions with the same sign of $\lambda$ are physically equivalent.
In our convention the solution with $\lambda < 0$ corresponds to the tachyon
rolling to the direction of the tachyon vacuum without D-branes.
The tachyon profile of the solution
as a function of time $x^0$
takes the following form~\cite{Schnabl:2007az,Kiermaier:2007ba}:
\begin{equation}\label{profile}
 T(x^0) =  \sum_{n=1}^\infty  \beta^{(n)}\,  \lambda^n \, e^{\frac{1}{\sqrt{\alpha'}} n x^0} \, . \end{equation}
The coefficients $\beta^{(n)}$ are
obtained by evaluating
\begin{equation}
\begin{split}
\beta^{(n)}\,  \lambda^n = \langle \, \phi^{(n)} , \Psi \, \rangle_{\rm density} \,,
\end{split}
\end{equation}
where
\begin{equation}
\phi^{(n)} (t)
= -c \partial c \, \phi_m^{(n)} (t) \,, \qquad
\phi_m^{(n)} (t)
= e^{-\frac{n}{\sqrt{\alpha'}} X^0} (t)\,.
\end{equation}
Here and in what follows the subscript
 `density' is used
to denote the quantity divided by the spacetime volume.
The weight $h$ of the operator $\phi_m^{(n)} (t)$ is given by $h=n^2$.

While the calculations of $\beta^{(n)}$
for the previous solutions~\cite{Schnabl:2007az,Kiermaier:2007ba,Kiermaier:2007vu}
were complicated,
the calculation of $\beta^{(n)}$ for (\ref{Psi2}) reduces to that of $C_\phi$.
A convenient way to calculate $C_\phi$ is to take the following limit:
\begin{equation}
\begin{split}
C_\phi & =
\bigl| \, \tfrac12(z+1)(z-1)\, \bigr|^{h} \,
\mla \, \phi_m (z) \,
\sigma_L (-1) \,
\sigma_R (1) \,
\mra_{\rm UHP} \\[.5ex]
& =
 2^{-h}
\lim_{z \to \infty} z^{2 h} \,
\mla \, \phi_m (z) \,
\sigma_L (-1) \,
\sigma_R (1) \,
\mra_{\rm UHP} \,.
\end{split}
\end{equation}
The three-point function
for $\phi_m^{(n)}$
is given by
\begin{equation}
\begin{split}
 \mla \, \phi_m^{(n)} (z) \,
\sigma_L (-1) \,
\sigma_R (1) \,
\mra_{\rm UHP}
&= \,\mla \, \phi_m^{(n)} (z) \,
\exp \, \biggl[ \, \lambda \int_{-1}^1 dt\,V(t) \, \biggr] \,
\mra_{\rm UHP} \\
& = \frac{\lambda^n}{n!}
\int_{-1}^1 dt_1 \int_{-1}^1 dt_2 \ldots \int_{-1}^1 dt_n \,
\mla \, \phi_m^{(n)} (z) \,
V(t_1) \, V(t_2) \ldots V(t_n) \,
\mra_{\rm UHP} \,.
\end{split}
\end{equation}
Since
\begin{equation}
\begin{split}
& \mla \, e^{-\frac{n}{\sqrt{\alpha'}}\,X^0}\!(z) \,
e^{\frac{1}{\sqrt{\alpha'}}\,X^0} \!(t_1) \,
e^{\frac{1}{\sqrt{\alpha'}}\,X^0} \!(t_2) \, \ldots
e^{\frac{1}{\sqrt{\alpha'}}\,X^0} \!(t_n) \,
\mra_{\rm UHP,\, density}
~=~
\prod_{i=1}^n \frac{1}{(z - t_i)^{2n}} \,
\prod_{i<j} (t_i - t_j)^2 \,,
\end{split}
\end{equation}
we obtain
\begin{equation}
C_{\phi^{(n)}, \, {\rm density}}
= \frac{\lambda^n}{2^{n^2} n!} \, I_n \,,
\end{equation}
where
\begin{equation}
I_n = \int_{-1}^1 dx_1 \int_{-1}^1 dx_2 \ldots \int_{-1}^1 dx_n
\prod_{i<j} (x_i - x_j)^2 \,.
\end{equation}
The integral $I_n$ is evaluated in appendix~\ref{secintegral},
 and we find
 \begin{equation}
    I_n=2^{n^2}n!
    \prod_{i=0}^{n-1}
    \frac{ i!^4}{(2i+1)!(2i)!}\,.
 \end{equation}
 It follows that
 \begin{equation}
 \myBoxed{
    C_{\phi^{(n)}, \, {\rm density}}
    =  \lambda^n
    \prod_{i=0}^{n-1}
    \frac{ i!^4}{(2i+1)!(2i)!}\,.
    }
 \end{equation}

We can now study the large-$n$ behavior of $\beta^{(n)}$
analytically. We use Sterling's approximation to find
\begin{equation}
    C_{\phi^{(n)}, \, {\rm density}}\,
    =\,\lambda^n\exp\Bigl[-2n^2\log 2  +{\cal O}(n\log n)\Bigr]\,.
\end{equation}
Combining this with
the asymptotic behavior~(\ref{asyg}) of $g(h)$,
we obtain
\begin{equation}
    \beta^{(n)}=
    \exp\Bigl[-\gamma\, n^2  +{\cal O}(n\log n)\Bigr]\,\qquad\text{with}~~
    \gamma\,=\,\log \frac{\pi}{2} \,.
\end{equation}
It is obvious from this exponential suppression
that the tachyon profile~(\ref{profile}) converges
at arbitrary time $x_0$.
While numerical fits
suggested convergence for
the solution in~\cite{Schnabl:2007az,Kiermaier:2007ba},
the current solution allows an analytic proof of this convergence.

The resulting profile is highly oscillatory at large $x_0$.
This feature of rolling tachyon solutions in string field theory
was first observed
in Siegel gauge by level truncation~\cite{Moeller:2002vx,Coletti:2005zj}
and later confirmed
in Schnabl gauge by the analytic solution~\cite{Schnabl:2007az,Kiermaier:2007ba}.
While this peculiar behavior of the rolling tachyon had been a puzzle in string field theory,
it was shown in~\cite{Kiermaier:2008qu} that the BCFT boundary state
for the rolling tachyon studied by Sen~\cite{Sen:2002nu} can be constructed from the solution
in string field theory.
It would be interesting
to see if one can
extract the closed string physics from the late-time behavior
of our solution.

The oscillatory behavior in the rolling tachyon profile is linked to the appearance of a suppression factor $e^{-\gamma n^2}$ in the coefficients $\beta^{(n)}$.
For the analytic solution~\cite{Schnabl:2007az,Kiermaier:2007ba} in Schnabl gauge,
the suppression factor
was not analytically determined but
numerically estimated in~\cite{Schnabl:2007az} as  $n^{-0.38 \, n^2}$\,.
For our solution based on bcc
operators,
we found
$e^{-\gamma n^2}$  with
$\gamma= \log(\pi/2)\approx 0.45$\,.
It is an interesting question if we can construct calculable analytic solutions without the dominant oscillatory
behavior, {\it i.e.},
solutions with $\gamma =0$.
Let us consider the value of $\gamma$ for solutions
associated with general projectors,
which can be generated from our solution based on the sliver projector
by reparameterizations~\cite{Okawa:2006sn}.
As we mentioned before, the large-$h$ behavior of $g(h)$ is determined by
considering $x=y=\frac{1}{2}$ in the limit $s\to\infty$ in~(\ref{gfunction}).
Combining this with~(\ref{h-dependence}), we find that $\gamma$ is determined from
\begin{equation}
\mla \, \sigma_R (-\tfrac{1}{2}) \,
f \!\circ\! \phi_m (0) \,
\sigma_L (\tfrac{1}{2}) \,
\mra_{{\rm UHP}}
= C_\phi \, \Bigl(\frac{8}{\pi}\Bigr)^h \,.
\end{equation}
For $\phi = \phi^{(n)}$, this reproduces $\gamma = \log (\pi/2)$ as follows:
\begin{equation}
C_{\phi^{(n)}, \, {\rm density}} \,
\Bigl(\frac{8}{\pi}\Bigr)^{n^2}\sim \lambda^n\Bigl(\frac{2}{\pi}\Bigr)^{n^2} =\, \lambda^n\,e^{-\gamma n^2} \,.
\end{equation}
We can use reparameterizations to generate new solutions,
and then the function $f(\xi)$ in~(\ref{arctan}) is replaced by
a general function $f(\xi)$ with $f(i) = \infty$~\cite{Okawa:2006sn}.
We usually choose $f(0)=0$ and
 $f(-1)=-f(1)$.
The large-$h$ behavior is then determined by
\begin{equation}
\mla \, \sigma_R (f(-1)) \,
f \!\circ\! \phi_m (0) \,
\sigma_L (f(1)) \,
\mra_{{\rm UHP}}
= C_\phi \, \biggl| \frac{f'(0) \, (f(1)-f(-1))}{(f(0)-f(1)) \, (f(0)-f(-1))} \biggr|^h
= C_\phi \, \biggl( \frac{2 f'(0)}{f(1)} \biggr)^h \,,
\end{equation}
and $\gamma$ is given by
\begin{equation}\label{gamma}
\gamma = \log \biggl( \frac{2 f(1)}{f'(0)} \biggr) \,.
\end{equation}
For a generic choice of $f(\xi)$, the tachyon profile is not calculable,
but it is calculable for solutions associated with special projectors~\cite{Rastelli:2006ap}.
A one-parameter family of special projectors labeled by $s$ with $s \ge 1$
 was introduced in~\cite{Rastelli:2006ap}. The associated function $f_s (\xi)$ and coefficient $\gamma_s$ are given by
\begin{equation}
f_s(\xi)\,=\,\xi \, {}_2F_1\bigl[\tfrac{s}{2},s;1+\tfrac{s}{2};-\xi^2\bigr]^{1/s} \,,
  \qquad~\gamma_s\,=\, \tfrac{1}{s}\log\!\frac{\sqrt{\pi}\,
  \Gamma \bigl(1+\tfrac{s}{2} \bigr)}{\Gamma \bigl( \tfrac{1}{2}+\tfrac{s}{2} \bigr)}\,.
\end{equation}
The sliver projector corresponds to $s=1$,
and the butterfly projector corresponds to $s=2$.
The  coefficient
 $\gamma_s$ for this family of solutions decreases as the parameter $s$ is increased.
However, one cannot choose $s$ such that the dominant oscillatory behavior is absent.
Indeed, $\gamma_s$
vanishes only in the limit $s \to \infty$, which corresponds to a singular limit of special projectors.

\section*{Acknowledgments}
We would like to thank Ian Ellwood, Ted Erler, Kazuo Hosomichi,  Michael Kroyter, Martin Schnabl,
\'Angel Uranga,
and Barton Zwiebach for valuable discussions.
M.K. and Y.O. would like to thank the organizers and participants of
the
KITP Workshop on {\em Fundamental Aspects of Superstring Theory},
the Simons Center for Geometry and Physics
 Workshop on {\em String Field Theory},
and the YITP workshop on  {\em Branes, Strings and Black Holes}
for stimulating discussions at various stages of this project.
Y.O. also thanks APCTP
for hospitality
during the Focus Program on {\em Current Trends in String Field Theory},
where a part of this work was presented.
P.S. thanks the DESY theory group for hospitality during a preliminary stage of this work.
The research of M.K. is supported by NSF grant PHY-0756966.
The work of
P.S. has been supported by the Spanish National Research Council (CSIC) JAE-Pre-0800401, and by Plan Nacional de Altas Energ\'\i as, FPA2009-07908,
Comunidad de Madrid HEPHACOS S2009/ESP-1473.
The work of Y.O. was supported in part
by Grant-in-Aid for Young Scientists~(B) No.~21740161 from
the Ministry of Education, Culture, Sports, Science and Technology (MEXT) of Japan
and by Grant-in-Aid for Scientific Research~(B) No.~20340048 from
the Japan Society for the Promotion of Science (JSPS).

\appendix
\section{Details of the derivation of $\Psi$}
\setcounter{equation}{0}

\subsection{Derivation of~(\ref{alternative})}\label{secaltderiv}
We now derive the alternative form~(\ref{alternative}) of the solution $\Psi$ starting from~(\ref{erler}).
It is useful to note that
\begin{equation}
\frac{1}{1 - h(K) \, BcV}
= 1 - Bc + \frac{1}{1 - h(K) \, V} \, Bc \,,
\end{equation}
where $h(K)$ is an arbitrary function of $K$.
This can be shown as follows:
\begin{equation}
\begin{split}
\frac{1}{1 - h(K) \, BcV}
& = 1 + h(K) \, BcV + h(K) \, BcV \, h(K) \, BcV + \ldots \\
& = 1 + h(K) \, V \, Bc + h(K) \, V \, h(K) \, V \, Bc + \ldots \\
& = 1 - Bc + \frac{1}{1 - h(K) \, V} \, Bc \,.
\end{split}
\end{equation}
It follows from
\begin{equation}
\biggl[ \, 1 - B \,
\frac{f(K)^2-1}{K} \, \lambda \, cV \, \biggr]^{-1}
= 1 - Bc + \biggl[ \, 1 - \frac{f(K)^2-1}{K} \,
\lambda \, V \, \biggr]^{-1} Bc
\end{equation}
and
$cV \, (1 - Bc) = 0$
that
\begin{equation}
\begin{split}\label{appPsi}
\Psi & = f(K) \, \lambda \, cV \,
\biggl[ \, 1 - \frac{f(K)^2-1}{K} \,
\lambda \,V \, \biggr]^{-1} Bc \, f(K) \,.
\end{split}
\end{equation}
This is precisely the form of $\Psi$ presented in~(\ref{alternative}).

\subsection{Derivation of $f(K) = 1 / \sqrt{1-K}$}\label{f^2=1/(1-K)}

Our starting point is the form~(\ref{appPsi}) of the solution $\Psi$.
We demand that the factor
\begin{equation}\label{factor}
\biggl[ \, 1 - \frac{f(K)^2-1}{K} \,
\lambda \,V \, \biggr]^{-1}
\end{equation}
can be written in terms of wedges with insertions of
finitely many
bcc operators. In particular,
we are interested in the case where
it can be written as a sum over terms of the form
\begin{equation}\label{goal}
    h_1(K+\lambda V)f_1(K) h_2(K+\lambda V)f_2(K) \ldots
    h_k(K+\lambda V)f_k(K)
\end{equation}
 with
the functions $h_i$ (with finite index range $i=1,\ldots,k
\leq\! k_{\rm max}
 \!<\!\infty$)
of $K +\lambda V$
in
the following form:
\begin{equation}
    h_i ( K+\lambda V )
    =\int_0^\infty d \alpha \, \tilde{h}_i (\alpha) \, \,e^{\alpha (K + \lambda V)}
    =\int_0^\infty d \alpha \, \tilde{h}_i (\alpha) \, \sigma_L\,e^{\alpha K}\,\sigma_R\,.
\end{equation}
To write the factor~(\ref{factor}) in the form~(\ref{goal}), we use $\lambda V=(K+\lambda V)-K$ and obtain
\begin{equation}
\biggl[ \, 1 - \frac{f(K)^2-1}{K} \,
\lambda \,V \, \biggr]^{-1}
\!\!=\biggl[ \, f(K)^2 - \frac{f(K)^2-1}{K} \,
(K+\lambda \,V) \, \biggr]^{-1}
\!\!=\sum_{k=0}^\infty \biggl[ \frac{f(K)^2-1}{Kf(K)^2} \,
(K+\lambda \,V) \, \biggr]^{k}\,f(K)^{-2}\,.
\end{equation}
Each term in this sum is of the general form~(\ref{goal}), with $h_i(K+\lambda V)=K+\lambda V$.
Unfortunately, this is singular because
\begin{equation}
K+\lambda V
= -\int_{-\infty}^\infty d \alpha \, \delta' (\alpha) \, \sigma_L\,e^{\alpha K}\,\sigma_R \,,
\end{equation}
which has support
at $\alpha=0$ only.
Consequently, all insertions of bcc operators collide, and no
finite-width wedges
with changed boundary conditions appear in the solution.
In addition, this form does not have the uniform bound $k_{\rm max}$
on the number of bcc operators.
However, if we have
\begin{equation}\label{fKcondition}
    \frac{f(K)^2-1}{Kf(K)^2}
    = a \,,
\end{equation}
for some constant $a$
independent of $K$, then
\begin{equation}
\begin{split}
\biggl[ \, 1 - \frac{f(K)^2-1}{K} \,
\lambda \,V \, \biggr]^{-1}
&=\sum_{k=0}^\infty \bigl[
\, a \,
(K+\lambda \,V) \, \bigr]^{k}\,f(K)^{-2}=
\frac{1}{1-
a \,
(K+\lambda V)}\,f(K)^{-2}\,.
\end{split}
\end{equation}
This expression is now a single term of the form~(\ref{goal}), with $k=1$\,,
$h_1(x)=1/(1-a x)$
and $f_1(x)=f(x)^{-2}$.
For $a>0$,
$h_1$ has the smooth Laplace transform
$\tilde{h}_1(\alpha)=a^{-1} e^{- \alpha/a}$,
which
vanishes at
$\alpha \to \infty$.
Thus~(\ref{fKcondition}), together with
$a>0$,
is the desired condition on $f(K)$.
Solving it for $f(K)$,
one obtains
\begin{equation}
    f(K)=\frac{1}{\sqrt{1-a K}}\,.
\end{equation}
In this case, we find $f_1(K) = 1-a K$, which is also acceptable.
We can use reparameterization~\cite{Okawa:2006sn}
to transform $K$, $B$, $c$, and $V$ as
\begin{equation}
K \to \beta K \,, \qquad
B \to \beta B \,, \qquad
c \to \frac{1}{\beta} \, c \,, \qquad
V \to \beta \, V \,.
\end{equation}
If we choose
$\beta = 1 / a$,
we have
\begin{equation}
f(K) = \frac{1}{\sqrt{1-K}} \,, \qquad
\Psi = \frac{1}{\sqrt{1-K}} \, \lambda \, cV \,
\biggl[ \, 1 - \frac{1}{1-K} \,
\lambda \, V \, \biggr]^{-1} Bc \,
\frac{1}{\sqrt{1-K}} \,.
\end{equation}
This form is thus unique up to reparameterization.

\section{Evaluation of the integral $I_n$}\label{secintegral}
\setcounter{equation}{0}

In this appendix we evaluate the following integral:\footnote{
We thank Kazuo Hosomichi for explaining the method in detail.
}
\begin{equation}\label{integral}
I_n = \int_{-1}^1 dx_1 \int_{-1}^1 dx_2 \ldots \int_{-1}^1 dx_n
\prod_{i<j} (x_i - x_j)^2 \,.
\end{equation}
The integral can be written as
\begin{equation}
I_n = \int_{-1}^1 dx_1 \int_{-1}^1 dx_2 \ldots \int_{-1}^1 dx_n \,
\Delta_n (x_1, x_2, \ldots, x_n)^2 \,,
\end{equation}
where
\begin{equation}
\begin{split}
\Delta_n (x_1, x_2, \ldots, x_n) & \equiv \left|
\begin{array}{cccccc}
1 & 1 & 1 & \ldots & 1 & 1 \\
x_1 & x_2 & x_3 & \ldots & x_{n-1} & x_n \\
x_1^2 & x_2^2 & x_3^2  & \ldots & x_{n-1}^2 & x_n^2 \\
\vdots & \vdots & \vdots & \ddots & \vdots & \vdots \\
x_1^{n-1} & x_2^{n-1} & x_3^{n-1}  & \ldots
& x_{n-1}^{n-1} & x_n^{n-1} \\
\end{array}
\right| \,.
\end{split}
\end{equation}
Let us rewrite $\Delta_n (x_1, x_2, \ldots, x_n)$
using the Legendre polynomials $P_n (x)$, which are given by
\begin{equation}
P_n (x) = \frac{1}{2^n n!} \, \frac{d^n}{dx^n} \, (x^2-1)^n\,,\qquad
\int_{-1}^1 dx \, P_n (x) \, P_m (x)= \frac{2}{2n+1} \, \delta_{nm} \,.
\end{equation}
The normalized polynomials $\widehat{P}_n (x)$ defined by
\begin{equation}
\widehat{P}_n (x) \equiv \frac{1}{c_n} P_n (x) \qquad\text{ with }~~~c_n = \frac{1}{2^n n!} \, \frac{(2n)!}{n!}\,
\end{equation}
have the form
$\widehat{P}_n (x) = x^n + \ldots$
and satisfy
\begin{equation}
\int_{-1}^1 dx \, \widehat{P}_n (x) \, \widehat{P}_m (x)
= a_n \, \delta_{nm}
\qquad\text{with }~~~a_n = \frac{2}{2n+1} \, \frac{1}{c_n^2} \,.
\end{equation}
Then the determinant $\Delta_n (x_1, x_2, \ldots, x_n)$ can be written as
\begin{equation}
\begin{split}
\Delta_n (x_1, x_2, \ldots, x_n) & = \left|
\begin{array}{cccccc}
\widehat{P}_0 (x_1) & \widehat{P}_0 (x_2) & \widehat{P}_0 (x_3)
& \ldots & \widehat{P}_0 (x_n) \\
\widehat{P}_1 (x_1) & \widehat{P}_1 (x_2) & \widehat{P}_1 (x_3)
& \ldots & \widehat{P}_1 (x_n) \\
\vdots & \vdots & \vdots & \ddots & \vdots \\
\widehat{P}_{n-1} (x_1) & \widehat{P}_{n-1} (x_2)
& \widehat{P}_{n-1} (x_3) & \ldots & \widehat{P}_{n-1} (x_n) \\
\end{array}
\right| \,,
\end{split}
\end{equation}
and we find
\begin{equation}\label{Inresult}
I_n = \int_{-1}^1 dx_1 \int_{-1}^1 dx_2 \ldots \int_{-1}^1 dx_n \,
\Delta_n (x_1, x_2, \ldots, x_n)^2
\,=\, n! \, \prod_{i=0}^{n-1} a_i \,=\,   2^{n^2}n!
\prod_{i=0}^{n-1}
\frac{ i!^4}{(2i+1)!(2i)!}\,.
\end{equation}
It is easy to verify this formula when $n=1, 2, 3$\,. An explicit evaluation of the integral~(\ref{integral}) gives
\begin{equation}
I_1\,=\,2\,,\qquad I_2 \,=\,\frac{8}{3}\,,\qquad I_3\,=\,\frac{64}{45}\,,
\end{equation}
in agreement with~(\ref{Inresult}).

\end{document}